\documentclass[twocolumn]{aastex701}
\usepackage[symbol]{footmisc}
\usepackage[T1]{fontenc}
\usepackage{subfigure}

\begin{document}

\newcommand{\water}{$\rm{H_2O}$ }
\newcommand{\mum}{$\rm{\mu m}$ }

\title{Up in smoke: carbon-destruction in the terrestrial planet-forming region of a young Class I disk}

\author[orcid=0009-0006-2641-5353,sname='Ballieux']{F.J. Ballieux }
\affiliation{Leiden Observatory, Leiden University, PO Box 9513, 2300 RA Leiden, The Netherlands}
\email[show]{ballieux@strw.leidenuniv.nl}  

\author[orcid=0000-0003-1878-327X, sname='McClure']{M.K. McClure} 
\affiliation{Leiden Observatory, Leiden University, PO Box 9513, 2300 RA Leiden, The Netherlands}
\email{mcclure@strw.leidenuniv.nl}

\author[orcid=0000-0003-4179-6394, sname='Bergin']{E.A. Bergin}
\affiliation{Department of Astronomy, University of Michigan, 1085 South University Avenue, Ann Arbor, MI 48109, USA}
\email{ebergin@umich.edu}

\author[orcid=0000-0002-2555-9869, sname='Van 't Hoff']{M.L.R. van 't Hoff}
\affiliation{Purdue University, Department of Physics and Astronomy, 525 Northwestern Avenue, West Lafayette, IN 47907, USA}
\email{vanthoff@purdue.edu}

\author[orcid=0000-0002-9470-2358, sname='Tychoniec']{Ł. Tychoniec}
\affiliation{Leiden Observatory, Leiden University, PO Box 9513, 2300 RA Leiden, The Netherlands}
\email{tychoniec@strw.leidenuniv.nl}

\author[orcid=0000-0001-8822-6327, sname='Francis']{L. Francis}
\affiliation{Leiden Observatory, Leiden University, PO Box 9513, 2300 RA Leiden, The Netherlands}
\email{francis@strw.leidenuniv.nl}

\author[orcid=0000-0001-7591-1907, sname='Van Dishoeck']{E.F. Van Dishoeck}
\affiliation{Leiden Observatory, Leiden University, PO Box 9513, 2300 RA Leiden, The Netherlands}
\affiliation{Max-Planck-Institut für Extraterrestrische Physik, Giessenbachstrasse 1, D-85748 Garching bei München, Germany}
\email{ewine@strw.leidenuniv.nl}

\begin{abstract}
Earth's surface and deep mantle are strongly depleted in carbon compared to the Sun. Sublimation of solid carbonaceous material in the inner midplanes of young disks could cause this depletion. We study such thermal processing in HOPS-68, a Class I protostar with a rich hydrocarbon spectrum. A companion paper has shown that it is recondensing refractory dust from a hot inner disk region, and that oxygen-bearing species are tracing a warm (300-500\,K) molecular layer entrained in a disk-wind.

In this work, we analyze the carbon species in the JWST spectrum of this source using local thermal equilibrium slab models. We detect warm ($\sim$400\,K) HCN and $\rm{C_2H_2}$ at the base of a disk wind, as well as colder ($\sim$40\,K) $\rm{C_4H_2}$, $\rm{CH_3}$, and CS in the outer disk or outflow cavity wall. Additionally, $\rm{^{13}CCH_2}$, $\rm{CH_4}$, and $\rm{C_6H_2}$ are detected, while ALMA observations reveal $\rm{C_2H}$ tracing the outflow cavity walls.

We suggest that in a "refractory carbon destruction zone" in the inner disk, carbon-bearing species are released from the grains, possibly contributing to the observed carbon depletion of Earth. The disk wind then redistributes carbon carried by HCN and $\rm{C_2H_2}$ from the inner disk to the colder outer disk, where UV-induced gas-phase reactions with $\rm{C_2H}$ in the outflow cavity walls could produce the aliphatic carbon chains seeb=n. 
This result implies that the Warm Carbon Chain Chemistry \citep[WCCC, e.g. ][]{sakai_warm_2013} might trace outflows rather than infall. The formation of aliphatic chains in the disk wind could promote the abundant aliphatic organics found in meteorites and asteroids.

\end{abstract}


\section{Introduction}
Carbon is the basis of all life as we know it; the dry mass of the human body consists of over 50\,\% carbon \citep{wang_estimation_2010}. However, the carbon abundance of the bulk silicate Earth (the entire Earth excluding the core) is depleted by four orders of magnitude compared to the Sun and the primordial dust in the interstellar medium (ISM) from which the Earth was born \citep{bergin_tracing_2015, bergin_carbon_2026}. 

There is no consensus yet on which mechanisms deplete carbon from the terrestrial planet-forming grains. Suggested mechanisms include photolysis and oxidation in the inner regions of protoplanetary disks 
\citep[e.g.][]{lee_solar_2010, anderson_destruction_2017, klarmann_radial_2018, vaikundaraman_refractory_2025}, or chemical sputtering at the surface of grains \citep{lenzuni_dust_1995, borderies_dust_2025}.
However, these mechanisms are insufficient to fully explain the depleted carbon content of Earth and small solar system bodies such as Ryugu \citep{nakamura_formation_2023}.  \cite{binkert_carbon_2023} find that, although oxidation and photolysis alone is not sufficient to explain the observed carbon depletion of Earth, increased vertical mixing and thermal processing during accretion outbursts could be an important contribution.

Because the refractory carbonaceous component of dust grains in the ISM consists of different components (i.e. pure amorphous carbon, aliphatic hydrocarbons, aromatic hydrocarbons, and graphite) which behave differently when heated \citep{gail_spatial_2017}, thermal processing will occur starting at temperatures of $\sim 300 - 500\,\rm{K}$ \citep[the `sootline', where moderately volatile carbon desorbs, ][]{lodders_jupiter_2004, kress_soot_2010,gail_spatial_2017,houge_burned_2025} increasing until even graphite sublimates. \cite{li_earths_2021} finds that thermal carbon depletion must already occur within the first million years of the Earth's formation. They suggest that in disks around young protostars, a $T>500\,\rm{K}$ region could reach radii of tens of AUs, through accretion heating.

In the inner few AUs of these young disks, the accretion-heated disk midplanes could even reach temperatures of 1400 K, where all dust, including the most refractory carbon-carrying species, sublimates in a so-called `thermostat region' \citep{cassen_utilitarian_1994, dalessio_effects_2006,mcclure_curved_2013, woitke_cai_2024}. We refer to this sublimation phase as a '$t=0$' moment, as it traces the start of the timeline of the Solar System's meteoritic record, when Calcium-Aluminum-rich Inclusions (CAIs) recondensed from hot gas in the disk midplane close to the proto-Sun.

Only two low-mass Class I protostars have been identified as candidates for the so-called $t=0$ moment: HOPS-315 \citep{mcclure_refractory_2025} and HOPS-68 \citep[][hereafter paper A]{ballieux_breath_2026}, both in the Orion Molecular Cloud (OMC) region. JWST spectra of both disks show strong gas phase absorption bands of warm $\rm{SiO}$, $\rm{H_2O}$, $\rm{CO_2}$, CO, and $\rm{^{13}CO}$. From the SiO we conclude that even the most refractory dust has begun to sublimate at the midplane in the inner few AU, releasing oxygen-rich molecules into the gas-phase. This oxygen-enhanced warm layer is contiguous with a disk wind. In both disks, silicate minerals are observed to have condensed out in the disk atmosphere, consistent with the $t=0$ condensation moment in our Solar System \citep{poteet_spitzer_2011,mcclure_refractory_2025}.

Both disks also show absorption bands of $\rm{C_2H_2}$ and HCN. In HOPS-315, these bands are shallow and seem to originate within the inner edge of the disk \citep{mcclure_refractory_2025}. However, in HOPS-68 they are much stronger and likely originate in the disk wind, similar to the SiO gas. 
The spectrum of HOPS-68 was also found to be extremely diverse in gas-phase hydrocarbon molecules, with the detection of absorption bands of $\rm{CH_3}$, $\rm{CH_4}$, $\rm{C_4H_2}$, $\rm{C_6H_2}$, CS, and $\rm{^{13}CCH_2}$. These species were not detected in HOPS-315. 

Diverse hydrocarbons dominate disks around very low-mass stars \citep[VLMS, e.g.][]{tabone_rich_2023, arabhavi_abundant_2024, arabhavi_minds_2025, kanwar_minds_2024, grant_transition_2025}, but are less pronounced around solar-mass protostars. Because of the differences in disk temperature, different chemical processes are likely at play in HOPS-68 relative to the VLMS disks. Notably, $\rm{C_6H_6}$ is commonly detected in VLMS sources, but was not detected in HOPS-68. Some higher-mass hydrocarbon-rich Class 0 and young Class I sources have been identified, with the detection of e.g. $\rm{C_4H_2}$ and $\rm{CH_4}$ \citep[e.g. ][]{van_gelder_jwst_2024, colmenares_jwstmiri_2024, colmenares_jwstmiri_2026}, however these molecular species are not common. In \citet{colmenares_jwstmiri_2026}, the small gas-phase hydrocarbons seen in the flat-spectrum disk ISO-Oph 37 are also entrained in a disk wind, suggesting that the removal of carbon-bearing molecules from the inner disk by winds could be a widely occurring phenomenon.

In paper A, we calculated the inner disk temperature structure of HOPS-68, showing that it has a 1400\,K thermostat region in its inner midplane. Therefore, in addition to moderately volatile carbon sublimation at the soot line (around 500\,K), HOPS-68 must experience higher-temperature processing of even the most refractory carbon components, like amorphous carbon, which does not thermally desorb until 1200\,K \citep{gail_spatial_2017}.
In this work, we aim to characterize HOPS-68's observed hydrocarbon species and use their properties to determine whether sublimation in the midplane could have contributed to the observed carbon depletion on Earth.

\section{Observations and methods}\label{sec:obs_meth}

\begin{figure*}[t!]
    \centering
    \includegraphics[width=\linewidth]{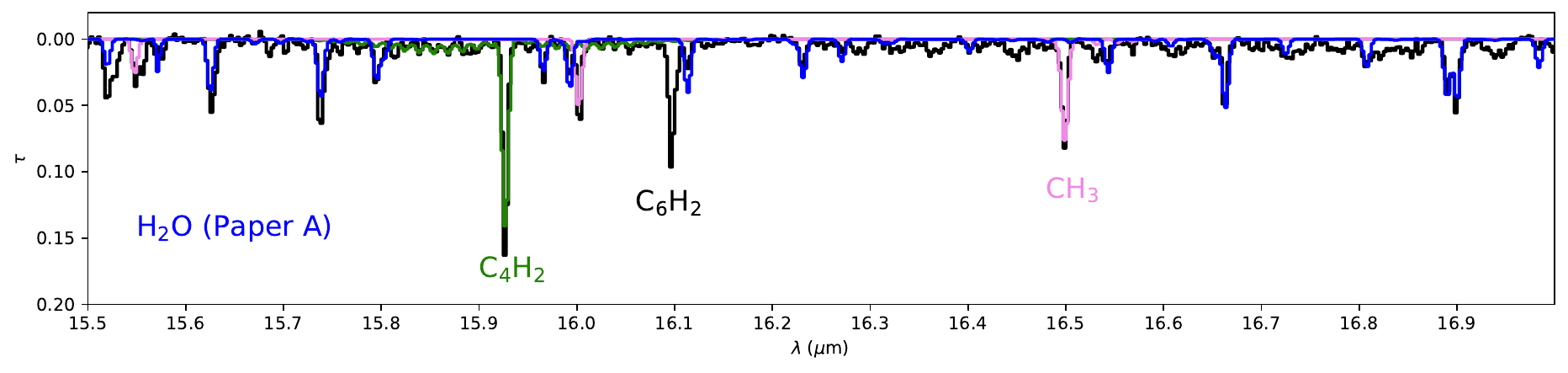}
    \caption{Slab model fits for $\rm{C_4H_2}$ and $\rm{CH_3}$ in the 15.5 - 17 \mum region, fit respectively to the residuals of the $\rm{H_2O}$ slab model from paper A. We also indicate the $\rm{C_6H_2}$ absorption band, to which we were unable to fit a slab model due to a lack of spectral data.}
    \label{fig:C4H2_etc_fit}
\end{figure*}

\begin{figure}[h!]
    \centering
    \includegraphics[width=\linewidth]{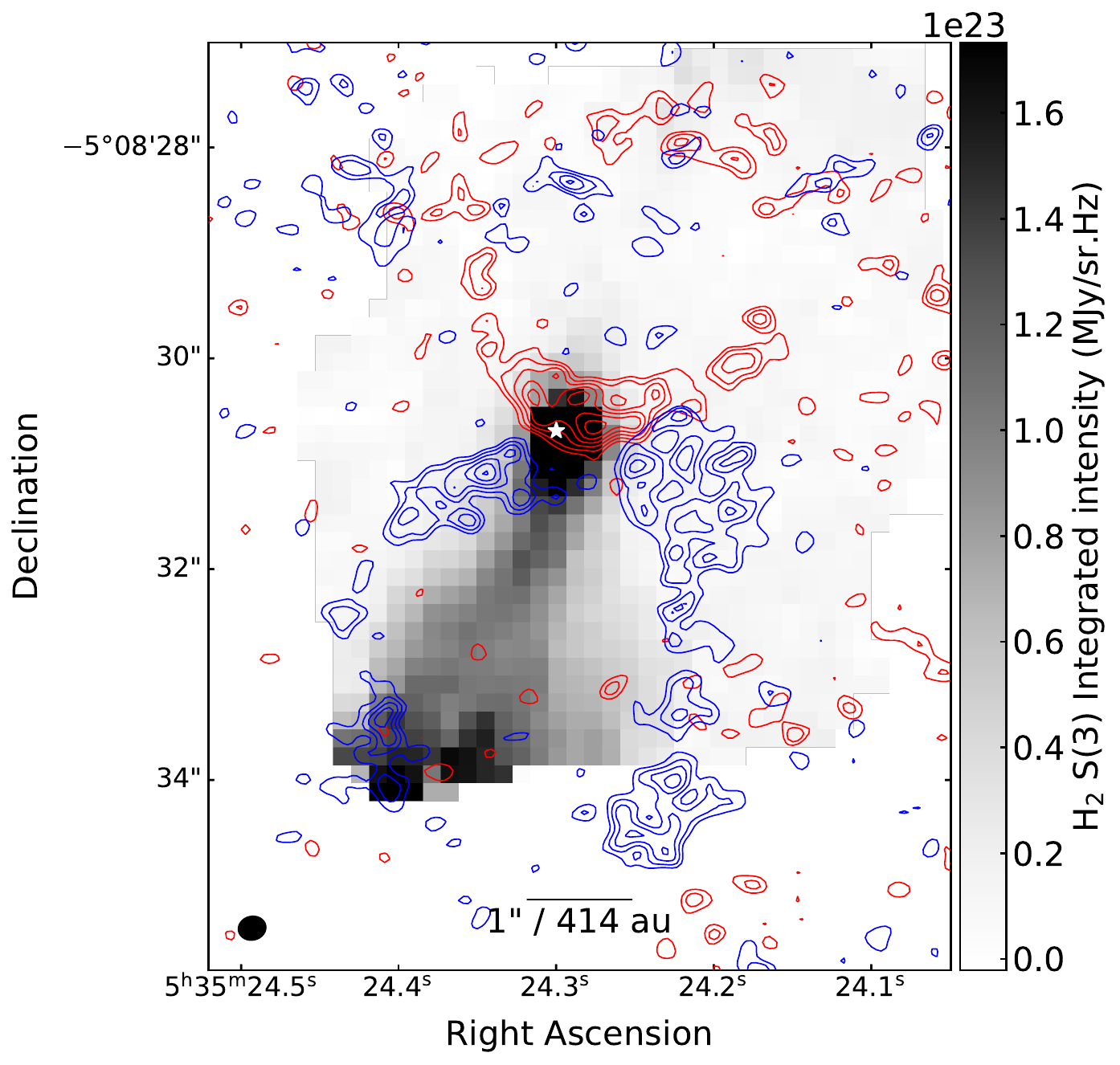}
    \caption{Integrated flux density of the JWST/MIRI $\rm{H_2\,S(3)}$ line at 9.6649 $\rm{\mu m}$, showing the blue-shifted $\rm{H_2}$ jet, and ALMA band 6 $\rm{C_2H}$ observations, centered at $\rm{262.065\,GHz}$, tracing the outflow cavity walls at $6 - 10\,\rm{km\, s^{-1}}$ (blue contours) and $12 - 20\,\rm{km\, s^{-1}}$ (red contours). Contours are plotted at 1 $\sigma$ intervals, starting at 2$\sigma$. 
    }
    \label{fig:ALMA_C2H}
\end{figure}

HOPS-68 is a $1.3\,\rm{L_\odot}$ protostar in OMC2 \citep[$414 \pm 7 \,\rm{pc}$][]{menten_distance_2007} inclined at 41\textdegree\ to our sightline \citep{poteet_spitzer_2011}, with a bolometric temperature of $T_{\rm{bol}}=100\,\rm{K}$, \citep[][]{tobin_vlaalma_2020} and a systematic velocity of $11.2\pm0.2\,\rm{km\,s^{-1}}$ (paper A). 
JWST observed HOPS-68 using the Mid-Infrared Instrument (MIRI) Medium Resolution Spectrometer \citep[$R\sim 2700-3300$, ][]{wright_mid-infrared_2023}, as part of the Cycle 1 General Observer program 1854, (\textit{It's COMplicated}, PI M.K. McClure\footnote{All the {\it JWST} data used in this paper can be found in MAST: \dataset[10.17909/1sj8-2b75]{http://dx.doi.org/10.17909/1sj8-2b75}.}). The full JWST/MIRI spectrum of HOPS-68 is shown in the appendix (Figure\,\ref{fig:full-spectrum}). 

We fitted local thermodynamic equilibrium (LTE) slab models to a number of absorption bands. A description of the observational setup, data reduction, and methods used in this work can be found in paper A, and is summarized in Appendix\,\ref{appendix:jwst_observations} and \ref{appendix:slabmodel}. The best-fit parameters for each molecular band fitted in this work are indicated in Table\,\ref{tab:full_results}. We show the best-fit slab models in Figure\,\ref{fig:C4H2_etc_fit} and in Figure\,\ref{fig:CH4_etc_fit} and \ref{fig:C2H2_HCN_fit} in the appendix.

Atacama Large (sub)-Millimeter Array (ALMA) band 6 observations of HOPS-68 were taken on the 27th of October 2016 (program 2016.1.00376.S, PI: López-Sepulcre), at an angular resolution of 0.23\,$"$ and a continuum sensitivity of 0.08\,$\rm{mJy\, beam^{-1}}$, targeting among others $\rm{C_2H}$, as shown in Figure\,\ref{fig:ALMA_C2H}. The primary beam corrected data cube was taken directly from the ALMA archive \footnote{https://almascience.nrao.edu/aq/} without further processing. More details can be found in Appendix\,\ref{appendix:ALMA_observations}.

\section{Results}\label{sec:results}

\begingroup
\renewcommand{\arraystretch}{1.4} 
\centering
\begin{table*}[htbp]
\fontsize{8pt}{8pt}\selectfont
\centering
\begin{tabular}{ll|lllll}
 Mole-            & $\lambda$          & $T$      & $N$            & $\Delta V$              & $v_{\rm{obs}}$ &$\sigma_\tau$\\ 
   cule          & [$\rm{\mu m}$]            & [K]    &[$\rm{cm^{-2}}$]& [km $\rm{s^-1}$]& [km $\rm{s^{-1}}$] &  \\ \hline \hline
HCN          & 12.5-14.5             &$319_{-24}^{+22}$  &  $4.5_{-0.7}^{+1.1} \cdot 10 ^{17}$  &  $9.5_{-1.0}^{+1.3}$  &  $-40.9_{-3.7}^{+3.6}$  &  0.04  \\
$\rm{C_2H_2}$& 12.5-14.5             &$432_{-16}^{+15}$  &  $1.1_{-0.1}^{+0.1} \cdot 10 ^{17}$  &  $3.5_{-0.2}^{+0.2}$  &  $-28.0_{-1.9}^{+1.9}$  &  0.01  \\ \hline
$\rm{C_2H_2}$&  7.3-7.8            & $451_{-93}^{+134}$  &  $2.3_{-0.5}^{+0.8} \cdot 10 ^{17}$  &  $[10]$  &  $[-28.0]$  &  0.02  \\ 
$\rm{CS}$&   7.3-7.8          & $54_{-14}^{+27}$  &  $2.8_{-1.0}^{+1.2} \cdot 10 ^{16}$  &  $[10]$  &  $[10.0]$  &  0.02  \\ \hline
$\rm{CH_3}$  & 15.5-17.0               &$44_{-7}^{+7}$  &  $8.0_{-0.8}^{+1.4} \cdot 10 ^{15}$  &  $21.7_{-15.0}^{+22.3}$  &  $7.3_{-5.6}^{+5.9}$  &  0.01       \\
$\rm{C_4H_2}$& 15.8-16.0             &$38_{-11}^{+12}$  &  $4.7_{-0.5}^{+0.9} \cdot 10 ^{15}$  &  $14.9_{-9.7}^{+12.6}$  &  $4.9_{-4.6}^{+4.6}$  &  0.01 \\ 
\end{tabular}
\caption{
The best-fit parameters from the slab model fits to different molecular bands. We indicate the wavelength range the models were fitted to, although the band sometimes extends beyond that range. The velocity shift $v_{\rm{obs}}$ is indicated in the LSRK frame. The source velocity of HOPS-68 in the LSRK frame is $v_{\rm{sys}}= 11.2\pm 0.2\,\rm{km\,s^{-1}}$ (paper A). The $\rm{CH_4}$ gas band is strongly influenced by an overlap with ice and other gas features, and is therefore not included in the table. We report $1\sigma$ uncertainties.}
\label{tab:full_results}
\end{table*}
\endgroup

\textbf{$\mathbf{C_2H_2}$, $\mathbf{HCN}$, and $\mathbf{^{13}CCH_2}$ -}
In the 12.5 - 14.5 \mum feature, a slab model for $\rm{C_2H_2}$ was fit to the residuals of the fit to the HCN band, the results of which are indicated in Figure\,\ref{fig:C2H2_HCN_fit}. 
The best-fitting slab model for HCN returns a temperature of $319_{-24}^{+22}$ K and a column density of  $4.5_{-0.7}^{+1.1} \times 10 ^{17}\,\rm{cm^{-2}}$. The $\rm{C_2H_2}$ is best fit by a slab model with a temperature of $432_{-16}^{+15}$ K and a column density of $1.1_{-0.1}^{+0.1} \times 10 ^{17}\,\rm{cm^{-2}}$. We also detect $\rm{^{13}CCH_2}$ at 13.73 \mum, however this band is too faint to obtain a reliable fit.

The $\rm{C_2H_2}$ band in the 7.3-7.8 \mum region, which was fit simultaneously with $\rm{CH_4}$ and CS, is best fit by a similar $T$ and $N$ compared to its 12.5-14.5 \mum counterpart. The linewidth of this band was fixed at $10\,\rm{km\,s^{-1}}$, based on typical values found in paper A. We fixed $v_{\rm{obs}}$ at $-28\,\rm{km\,s^{-1}}$ in the LSRK frame, to match the other $\rm{C_2H_2}$ band.

\textbf{CS -}
\label{sec:CS_results}
The CS feature in the 7.3-7.8 \mum region returns a value of $T=54_{-14}^{+27}\,\rm{K}$, with a column density of $2.8_{-1.0}^{+1.2} \times 10 ^{16}\,\rm{cm^{-2}}$. The linewidth of this band was fixed at $10\,\rm{km\,s^{-1}}$ and $v_{\rm{obs}}$ was fixed at $+10\,\rm{km\,s^{-1}}$ in the LSRK frame, which is close to the source velocity.

\textbf{$\mathbf{CH_4}$ -}
\label{sec:CH4_results}
The $\rm{CH_4}$ band in the 7.3-7.8 \mum region is best fit by a $\sim 1000\,\rm{K}$ temperature and a red-shifted velocity of $\sim 20 \,\rm{km\, s^{-1}}$. The linewidth of this band was fixed at $10\,\rm{km\,s^{-1}}$, based on typical values found in other bands. However the quality of the fit is limited by overlap with other gas bands and the $\rm{CH_4}$ ice band at the same wavelength range. Varying the continuum subtraction allowed us to fit a large range of temperatures (between $\sim$ 10 and 1000\,K) to the band, indicating that the fitting results from this band are not robust. Therefore we do not consider this band further in this work. 

\textbf{$\mathbf{CH_3}$, $\mathbf{C_4H_2}$ and $\mathbf{C_6H_2}$ -}
\label{sec:16muhydrocarbon_results}
In the 15.5-17 \mum region, we fit $\rm{CH_3}$ and $\rm{C_4H_2}$ to the residuals subtracting the \water lines, see Figure\,\ref{fig:C4H2_etc_fit}.  
With temperatures of $44_{-7}^{+7}\,\rm{K}$ and $38_{-11}^{+12}\,\rm{K}$ respectively, these hydrocarbons are relatively cold. The column densities, $\Delta V$, and $v_{obs}$ of these species remain hard to constrain, as we only detected a narrow Q-branch, which consists of several blended lines, for both species. These species are likely at rest with the source velocity, although the uncertainties are relatively large. 

In the same region, we detect $\rm{C_6H_2}$ at 16.1 $\rm{\mu m}$ \citep{haas_high-resolution_1994}. Since $\rm{C_6H_2}$ is not in the LAMDA and HITRAN databases, we are unable to fit the band with the LTE slab model fits. Because we detect only a strong and narrow Q-branch of this species, similar to $\rm{C_4H_2}$ and $\rm{CH_3}$, we suspect the $\rm{C_6H_2}$ gas is likely relatively cold as well.

\textbf{ALMA $\mathbf{C_2H}$ observations -}
Figure\,\ref{fig:ALMA_C2H} presents the integrated flux density maps of the ALMA $\rm{C_2H}$ observations, integrated between 6 - 10 $\rm{km\,s^{-1}}$ in the red contours, and between 12 - 20 $\rm{km\,s^{-1}}$ in the blue contours. These clearly bracket the JWST/MIRI observations of the $\rm{H_2\,S(3)}$ jet, indicating that $\rm{C_2H}$ traces a wider component than the jet, likely the outflow cavity walls. Furthermore, the blue-shift of the $\rm{C_2H}$ emission follows that of the jet, indicating that it traces outflowing, rather than infalling, material. 

\section{Discussion} \label{sec:discussion}

\begin{figure}
        \centering
            \includegraphics[width=0.49\textwidth]{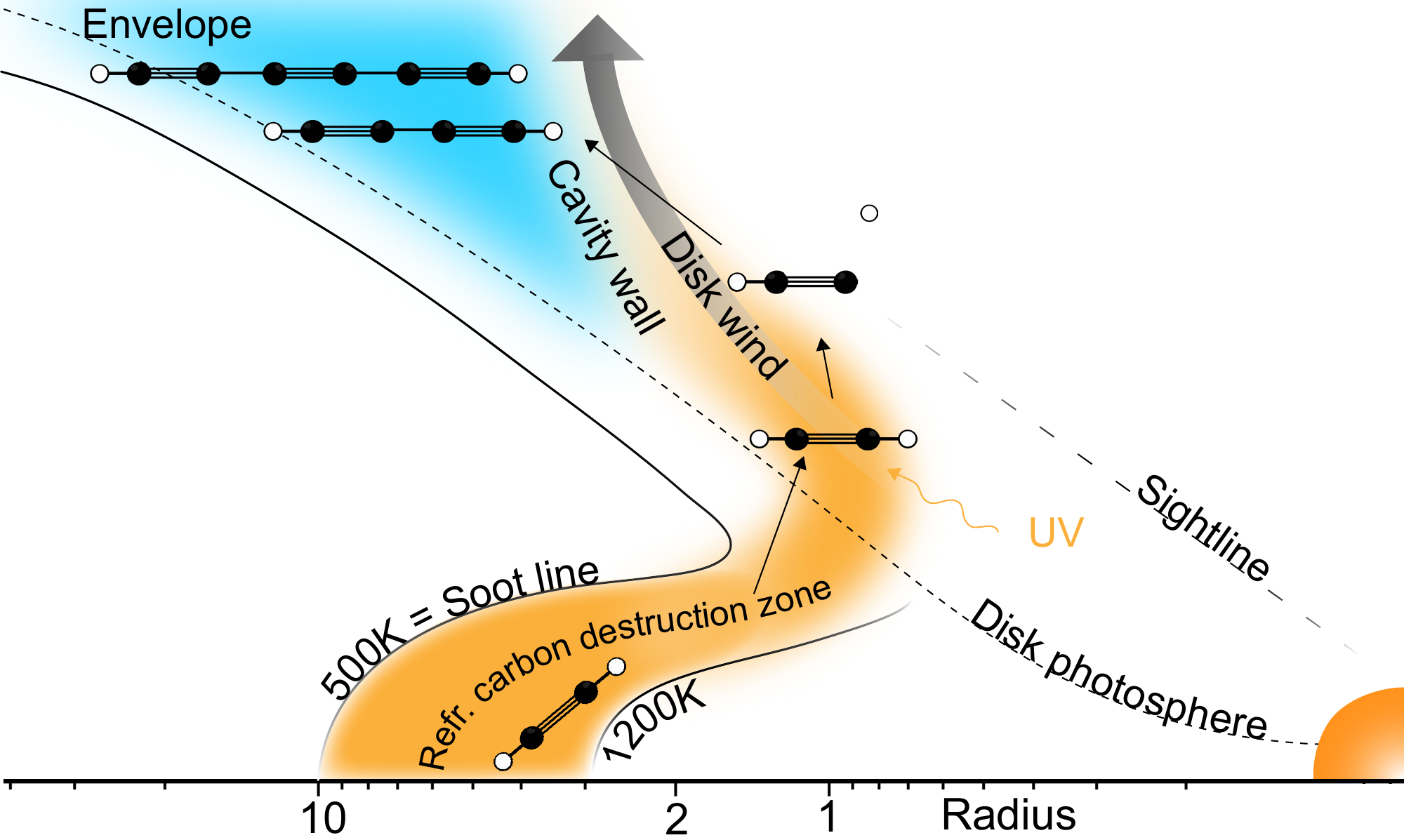}
        \caption{
        Proposed scenario to explain the hydrocarbon-rich spectrum of HOPS-68. Gas-phase carbon-bearing species enrich the gas between the soot line ($\sim 500\,\rm{K}$) and the graphite sublimation rim ($\sim 1200\,\rm{K}$), which we refer to as the "refractory carbon destruction zone", and are then entrained and transported out of the disk by the disk wind. In the outflow cavity, $\rm{C_2H_2}$ reacts with UV-photons, with $\rm{C_2H}$ as reaction product. $\rm{C_2H}$ can react with $\rm{C_2H_2}$ to form carbon-chain species such as $\rm{C_4H_2}$ and $\rm{C_6H_2}$. 
        The location of the isotherms were estimated using the results from paper A.} 
        \label{fig:diagram}
\end{figure}

\subsection{Carbon-bearing species detected}\label{sec:radii_here}
In paper A, HOPS-68's infrared bands of $\rm{H_2O}$, SiO, $\rm{CO_2}$ and $\rm{^{13}CO}$ were found to trace a warm molecular layer above the disk midplane, which ALMA SiO observations confirmed kinematically to be the base of a disk wind. 
In this work, we find high column densities of HCN and $\rm{C_2H_2}$ (a few times $10^{17}\,\rm{cm^{-2}}$), along with $\rm{^{13}CCH_2}$. These species were only detected in 2, 3, and 2 out of 18 low-mass protostars in the JOYS sample, respectively \citep{van_gelder_jwst_2024}. The $\rm{C_2H_2}$ and HCN column densities are approximately 20 times larger than detected in the Class I protostar HOPS-315 \citep{mcclure_refractory_2025}. 

The temperatures of HCN and $\rm{C_2H_2}$ (between $319$ and $432\,\rm{K}$) and blue-shifted velocities (by between $53$ and $39\,\rm{km\,s^{-1}}$ with respect to the source velocity), are similar to those found in paper A for the oxygen-bearing species, so we conclude that these hydrocarbons are in the same molecular layer.

Aside from these warm carbon-bearing species, we also detect CS, $\rm{CH_3}$, and $\rm{C_4H_2}$. Based on their $\sim$ 40 K temperatures, they are likely located at large radii, and their velocity shift we find they could be associated either with lower-velocity material at the outflow cavity walls or with the envelope. Since they are detected in absorption, they must be located above the hot disk photosphere, which provides the continuum against which they absorb. We therefore suggest that they are located in the upper layers of the outer disk or in the outflow cavity walls. In addition to these simple hydrocarbons, we have also detected triacetylene ($\rm{C_6H_2}$) for the first time in a protostellar system, via its 16.1\,\mum band. More information on this species can be found in Appendix\,\ref{appendix:carbonchains}.

\subsection{Origin of simple hydrocarbons}\label{dis:origin_hydrocarbons}
To understand the origin of the carbon-bearing species in HOPS-68, we first consider the simpler species entrained in the disk wind, namely $\rm{C_2H_2}$ and $\rm{HCN}$. These species could be released into the gas phase through the sublimation of moderately volatile carbon from grains, beginning at the soot line ($\sim 500\,\rm{K}$), after which they could be caught in the disk wind. However, an amorphous carbon component can remain on the grains until it, too, sublimates around 1200\,K \citep{gail_spatial_2017,lenzuni_dust_1995, borderies_dust_2025}. Therefore, in the disk region between 500 and 1200\,K, all the refractory carbon desorbs from the grains, and we refer to this region as the "refractory carbon destruction zone". We show the proposed scenario in Figure\,\ref{fig:diagram}. 

The thermal processing of carbon-bearing grains in the midplane could contribute to a depletion of carbon in terrestrial planet-forming regions, as the sublimated carbon-bearing molecules could either accrete onto the star or be blown out in the disk wind. Such thermal processing in a carbon destruction zone might also explain the observed carbon-depletion of Earth \citep[as also suggested by ][]{colmenares_jwstmiri_2026}. In HOPS-68, this could be amplified by the thermostat region (paper A) in the inner midplane pushing out and enlarging the carbon destruction zone.

To constrain the amount of carbon carried off in the wind, and to estimate its contribution to the abundances observed on Earth, we would have to quantitatively constrain the amount of gas in the disk wind, using e.g. the IR $\rm{H_2}$ lines or mm CO observations. Such an analysis is outside of the scope of this letter, but will be done in future work. 

Due to the faintness of these bands in HOPS-315, the velocity for the HCN and $\rm{C_2H_2}$ bands could not be determined, so it is unclear whether they are also wind-born in that source. However, our proposed scenario aligns with the conclusions of \cite{colmenares_jwstmiri_2024, colmenares_jwstmiri_2026}, who suggest that dust processing at the soot line explains high column densities of carbon-bearing species in some flat-spectrum and Class II disks.

\subsection{Carbon-chain chemistry in the outflow cavity}
The longer carbon-chain species are likely formed through a different mechanism than the smaller warmer hydrocarbons.
A fraction of the $\rm{C_4H_2}$ could be produced through the photodissociation of PAHs \citep[e.g. pyrene, ][]{marciniak_photodissociation_2021}, however we introduce an alternative formation route, that also accounts for the detection of $\rm{C_6H_2}$ through gas-phase chemistry along the outflow cavity walls. $\rm{C_2H}$ is the main photodissociation product of $\rm{C_2H_2}$ \citep[e.g. ][]{heays_photodissociation_2017} and can form through:

\begin{equation}
    \rm{C_2H_2 + h\nu \rightarrow C_2H + H}.
\end{equation}

Longer carbon-chains can then efficiently form through gas-phase reactions with $\rm{C_2H}$, as we describe in detail in Appendix\,\ref{appendix:carbonchains}. 
In HOPS-68, we detect $\rm{C_2H}$ outflowing along the outflow cavity walls with ALMA (Figure\,\ref{fig:ALMA_C2H}). $\rm{C_2H}$ is often associated with outflow cavities
\citep[e.g. ][]{oya_substellar-mass_2014, tychoniec_which_2021, anderson_peaches_2025, zhang_rotation_2018}.

\subsection{Carbon combustion in the refractory carbon destruction zone}
Between 900 and 1100\,K, PAH formation through carbon combustion could occur if there is a sufficient reservoir of gas-phase $\rm{C_2H_2}$ available \citep{frenklach_formation_1989, gail_radial_2001, gail_radial_2002}. These conditions are met in the 'refractory carbon destruction zone', and therefore an additional carrier of carbon in the disk wind could be PAHs formed through combustion in the carbon destruction zone. More modeling of carbon combustion would have to be done to confirm this.

The presence of PAHs in disks wind has been confirmed by JWST observations of edge-on T-Tauri disks \citep{arulanantham_jwst_2024, sturm_jwstmiri_2024, dartois_edge-disc_2025}. We do not detect PAH emission in the on-source JWST spectrum of HOPS-68; however a more advanced analysis of the off-source extended emission should be done to establish whether PAHs are present in the disk wind.

In the disk wind, such PAHs could release warm simple hydrocarbons such as $\rm{C_2H_2}$ into the gas-phase through the photodissociation. Laboratory experiments find that photodissociation of PAHs can produce a large range of hydrocarbon species, including $\rm{C_2H_2}$ \citep[the main photo-product,][]{murga_impact_2020, omont_carbon_2025}, $\rm{CH_4}$, and $\rm{CH_3}$ \citep{alata_vacuum_2015,diedhiou_structure_2019, marciniak_photodissociation_2021}, thereby possibly contributing to the abundances of the gas-phase carbon-bearing species detected in HOPS-68.

\subsection{Impact on protostar \& Solar System chemistry}
\textbf{Warm carbon-chain chemistry -} Carbon chain species have been found to be abundant at $\sim 25\,\rm{K}$ towards some low-mass protostars in (sub-)mm observations, e.g. L1527 \citep{sakai_abundant_2008} and IRAS 15398-3359 \citep{sakai_discovery_2009}.
A mechanism often invoked to explain them is Warm Carbon Chain Chemistry (WCCC), in which carbon chain molecules are efficiently produced from $\rm{CH_4}$ that has sublimated from icy grains in infalling envelope at 25\,K\citep{collings_laboratory_2004}. \footnote{It is important to note that in the context of protostellar cores, $25\,\rm{K}$ is considered to be warm, hence WCCC, while in the context of a more evolved protostar 25\,K would be considered cold. In this work, if we refer to warm temperatures, we refer to temperatures of several 100\,K.}

The WCCC mechanism is suggested to be the result of by material infalling from the envelope onto the disk \citep[e.g. ][]{aikawa_chemical_2020}. In contrast, the infrared carbon-bearing species in HOPS-68 are clearly associated with outflowing material in the disk wind, consistent with the results of the millimeter PEACHES survey of a much larger sample of protostars \citep{anderson_peaches_2025}. Infrared observations by \cite{mcclure_refractory_2025} and \cite{colmenares_jwstmiri_2026} also find $\rm{C_2H_2}$ and HCN entrained in a disk wind. A disk wind origin was first suggested to explain blue-shifted HCN and $\rm{C_2H_2}$ absorption bands in the Class I source Oph IRS46 \citep{lahuis_hot_2006}. The combination of the carbon destruction zone and the disk wind is the best mechanism to explain the formation of long hydrocarbon chains and PAHs where they are seen with JWST and ALMA, and might therefore be an alternative to WCCC. WCCC being related to outflows rather than infall is further supported by numerous hydrocarbons tracing the outflow cavity wall of WCCC source L1527 \citep{hoff_chemical_2026}.

It should be noted that, in addition to the already available ALMA detection of $\rm{C_2H}$ along the outflow cavity walls and the JWST detection of carbon chains, (sub-)mm observations of longer carbon chains in HOPS-68 are required to definitively confirm it as a WCCC source.

\textbf{Carbon and oxygen enrichment -}
In paper A, the presence of SiO gas as well as high abundances of $\rm{H_2O}$ showed that the gas in the inner regions of HOPS-68 must be enriched in oxygen, as a natural consequence of the thermostat region. In this work, we find high column densities of gas-phase carbon-bearing species, and suggest that this could be caused by carbon being released off the grains in a 'refractory carbon destruction zone'. If the excess oxygen and carbon are roughly balanced and in the same regions, we would expect them to combine into CO gas. However, in the spectrum of HOPS-68, we still find a rich variety of both oxygen- and carbon-bearing gas-phase molecules. Although there is abundant CO detected as well (paper A), we suggest that this diversity of molecules can survive because the oxygen and carbon enriched layers are launched into the disk wind at slightly different radii, and these layers do not mix. This is consistent with recent MHD wind and jet simulations, which show laminar wind flow along segregated streamlines at radii outside of the turbulent region surrounding the jet \citep{panoglou_molecule_2012,tu_modeling_2025}. We hypothesize that there is a layer of CO gas with a broad range of radii and temperatures at the interface of the oxygen-rich and carbon-rich disk wind layers, which could explain why the CO band could not be modeled by simple LTE slab models in paper A. The impact on the disk wind of the segregated oxygen- and carbon-enhancement provided by the thermostat region and the carbon destruction zone could contribute to the onion-like outflow structures across different stages of star formation \citep[e.g.][]{ray_outflows_2023,tychoniec_jwst_2024,delabrosse_jwst_2024,arulanantham_jwst_2024, pascucci_nested_2025}.

Furthermore, the variety of carbon-bearing species could persist in oxygen-rich gas due to self-shielding of \water, which locks up oxygen before it can react with carbon atoms \citep{duval_water_2022}. However, such a mechanism would suppress the abundance of CO and $\rm{CO_2}$. 

\textbf{HOPS-68 as Rosetta's stone for the meteoritic record -}
In paper A, we found that HOPS-68 \citep[like HOPS-315,][]{mcclure_refractory_2025} is undergoing the $t=0$ moment of the Earth meteoritic record, the phase where CAIs and amoeboid olivine aggregates (AOAs) recondense from a hot thermostat region in the inner disk. These CAIs and AOAs are found in meteorites, where they are embedded in a fine-grained matrix material consisting of mineral and carbonaceous phases mainly from fine dust present in the solar nebula, as well as trace amounts of presolar grains \citep{nittler_astrophysics_2016}, 
indicative of extensive radial mixing in the solar nebula. 

Sample return missions of carbonaceous asteroids Ryugu \citep{naraoka_soluble_2023, ito_pristine_2022} and Bennu \citep{glavin_abundant_2025} have revealed the presence of abundant aliphatic-rich organics, consistent with the matrix material of CI chondrite materials. We suggest that the hydrocarbon chain species detected in HOPS-68, $\rm{C_4H_2}$ and $\rm{C_6H_2}$, might contribute to the formation of these aliphatic organics. Since these species formed in the outflow cavity walls, they are not entrained in the disk wind, and thus will not leave the system like the lighter hydrocarbon species. They may be expected to fall into the outer disk at later times, consistent with accretion timescales for these two asteroids.

If we indeed detect the origin of some of the matrix material found in the meteoritic record, then HOPS-68 is truly Rosetta's stone for early planetesimal formation, where we simultaneously detect the formation of CAIs and AOAs (paper A) and the amorphous matrix material that these inclusions are caught in. 

\section{Conclusions}\label{sec:conclusions}
Earth is severely depleted in carbon, which years of modeling effort have not been able to account for. \cite{li_earths_2021} find that sublimation in the inner disks of very young protostars could play an important role in explaining this carbon depletion. Recently, work by \cite{mcclure_refractory_2025} and \cite{colmenares_jwstmiri_2026} have found evidence that gas-phase carbon-bearing species might be present in a disk wind.

In paper A, we presented the JWST spectrum of HOPS-68, a Class I protostar that hosts a 1400\,K thermostat region in its inner disk, as well as a warm (300-500\,K) warm molecular layer that is entrained in a disk wind, as traced by oxygen-bearing species. The JWST spectrum of HOPS-68 is also rich in gas-phase hydrocarbons, which are studied in this work. 

\begin{itemize}
    \item We use LTE slab models to constrain the gas temperature, column density and velocities of the gas traced by bands of $\rm{C_2H_2}$, HCN, $\rm{C_4H_2}$, $\rm{CH_3}$, $\rm{CH_4}$, and CS in the JWST spectrum. We further detect $\rm{C_6H_2}$ and $\rm{^{13}CCH_2}$. 
    \item We suggest that the warm simple hydrocarbons such as $\rm{C_2H_2}$ sublimate from dust grains in a "refractory grain destruction zone" in the inner disk, and are subsequently entrained in the disk wind. 
    Carbon sublimation in the inner disk of low-mass protostar HOPS-68 likely contributes to the mechanism responsible for the carbon-depletion of the Earth. 
    \item $\rm{C_4H_2}$, $\rm{CH_3}$, and CS are colder, and are likely present in the outer disk or extended envelope. $\rm{C_4H_2}$ and $\rm{C_6H_2}$ could form in the outflow cavity walls via chemistry induced by UV photons. This is supported by an ALMA detection of $\rm{C_2H}$ outflowing along the outflow cavity walls, indicative of $\rm{C_2H_2}$ photodissociation. 
    \item In the 'refractory carbon destruction zone', the conditions are likely met to allow for PAH formation through combustion, although a more detailed study is required to confirm the presence of PAHs in the disk wind of HOPS-68.
    \item We suggest that abundant carbon-chains in protostars, previously attributed to WCCC \citep[e.g.][]{sakai_warm_2013}, are formed in outflows rather than infalling envelopes. We further propose that the observed hydrocarbon chains in HOPS-68 might be an important contributor to the origin of the abundant aliphatic organic material found in the matrix of meteorites and asteroids such as Bennu and Ryugu.
\end{itemize}

To determine whether the observed carbon chain species are indeed formed through UV-induced chemistry in the outer disk or envelope, we would need larger-scale (sub)-mm observations of carbon-chain species, to see if their location coincides with the observed $\rm{C_2H}$ emission. Although pure rotational transitions of $\rm{C_4H_2}$ and $\rm{C_6H_2}$ would not be observable in the (sub)-millimeter range, other non-symmetric carbon chain species might serve as a proxy. Ground based IR and sub-mm observations at high spectral resolution might also help us constrain the kinematics of hydrocarbons further, revealing whether this mechanism can explain carbon abundance patterns in our solar system. Such observations are also required to definitively classify HOPS-68 as a WCCC source.

Future work will focus on constraining the amount of $\rm{H_2}$ in the disk wind and the jet, such that we can constrain the amount of carbon lost through the disk wind. 

A carbon-rich spectrum such as in HOPS-68 is not commonly found in Class I protostars. We likely observed HOPS-68 at exactly the right time, at exactly the right inclination angle to allow for these detections. Enlarging our sample of $t=0$ sources and monitoring the sources already identified would allow us to constrain how the spectrum changes over time, and what the influence of variability could be on the detection of these hydrocarbons. 

\begin{acknowledgements}
This work is based in part on observations made with the NASA/ESA/CSA JWST. The data were obtained from the Mikulski Archive for Space Telescopes (MAST) at the Space Telescope Science Institute, which is operated by the Association of Universities for Research in Astronomy, Inc., under NASA contract NAS 5-03127 for JWST. These observations are associated with programme no. 1854.This paper makes use of the following ALMA data: ADSJAO.ALMA\#2016.1.00376.S ALMA is a partnership of the ESO (representing its member states), NSF (USA) and NINS (Japan), together with NRC (Canada), MOST and ASIAA (Taiwan) and KASI (Republic of Korea), in cooperation with the Republic of Chile. The Joint ALMA Observatory is operated by ESO, auI/NRAO and NAOJ. The National Radio Astronomy Observatory and Green Bank Observatory are facilities of the U.S. National Science Foundation operated under cooperative agreement by Associated Universities, Inc.
F.J.B. and M.K.M were financially supported by the NWO Talent Programme Vidi (grant number 233.122) from the Dutch Research Council (NWO).
Ł.T. acknowledges support from the COST Action CA22133 PLANETS.
E.A.B. acknowledges support from NASA grant 80NSSC24K0149, 80NSSC20K0259, and NSF Award No.2509735.
E.F.v.D. and L.F. are supported by funding from the European Research Council (ERC) under the European Union’s Horizon 2020 research and innovation programme (grant agreement No. 101019751
MOLDISK) and by the Danish National Research Foundation through the Center of Excellence “InterCat” (Grant agreement no.:DNRF150).
M.L.R.H. acknowledges support from NSF Award No. 2509734

\end{acknowledgements}

\bibliography{bibliography}{}
\bibliographystyle{aasjournalv7}

\appendix

\begin{figure*}[h!]
    \centering
    \includegraphics[width=\linewidth]{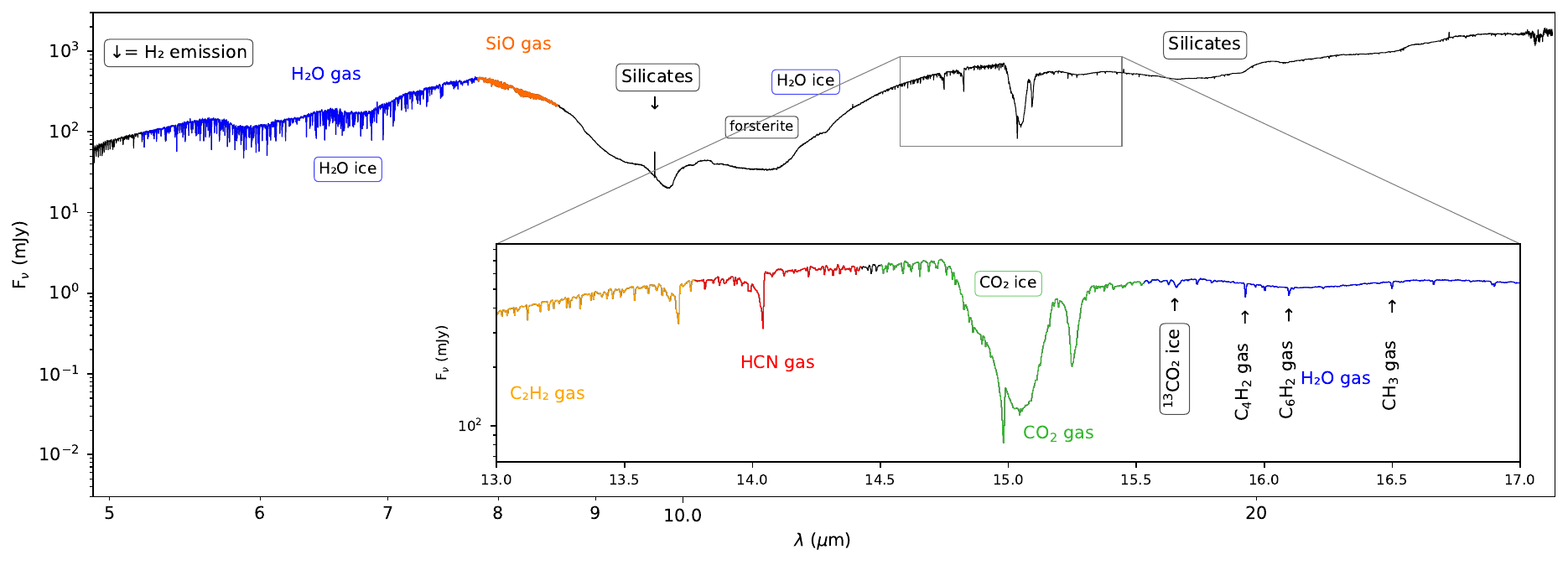}
    \caption{The full JWST/MIRI spectrum of HOPS-68, indicating the molecular gas absorption bands studied in this work and in paper A, as well as the most notable ice and silicate absorption features. Not indicated are the detections of $\rm{^{13}CCH_2}$, CS, and $\rm{CH_4}$ in the 7.3-7.8 \mum region. Some prominent $\rm{H_2}$ emission lines are highlighted with a black arrow. The inset shows a zoom-in to the 13–17 \mum region, which shows a number of notable hydrocarbon bands. SiO, \water, and $\rm{CO_2}$ bands are analyzed in detail in paper A.}
    \label{fig:full-spectrum}
\end{figure*}

\section{JWST Observations}\label{appendix:jwst_observations}
JWST MIRI observations were taken on the 16th of March 2023 for a total exposure time of 222 seconds. HOPS-68 is spatially unresolved over the entire JWST/MIRI wavelength range, except for a jet visible in $\rm{H_2}$ emission. The spectrum was extracted over a circular aperture with a diameter three times the full width at half maximum (FWHM) of the point-spread function (PSF) of the instrument. We perform all model fitting in the barycentric frame (the default for JWST), and convert the fitted velocities to the kinematic local standard of rest (LSRK) frame after the slab model fitting. Full details on the data reduction process can be found in paper A.

\section{ALMA observations}\label{appendix:ALMA_observations}
Primary beam corrected datacubes were taken from the ALMA science archive without further processing. There are a number of $\rm{C_2H}$ transitions in the observed frequency range, the most relevant ones being the ($N =3-2, J=\frac{5}{2}, F=3-2$) transition at 262.065\,GHz and the ($N =3-2, J=\frac{5}{2}, F=2-1$) at 262.067\,GHz. When extracting an aperture further from the central protostar, we can separate the two transitions, however on-source there is significant blending of the red- and blue-shifted lobes. To create the integrated velocity maps, we center our observations on the 262.065\,GHz transition, and select a velocity range that covers both transitions.

\begin{figure}[h!]
    \centering
    \includegraphics[width=\linewidth]{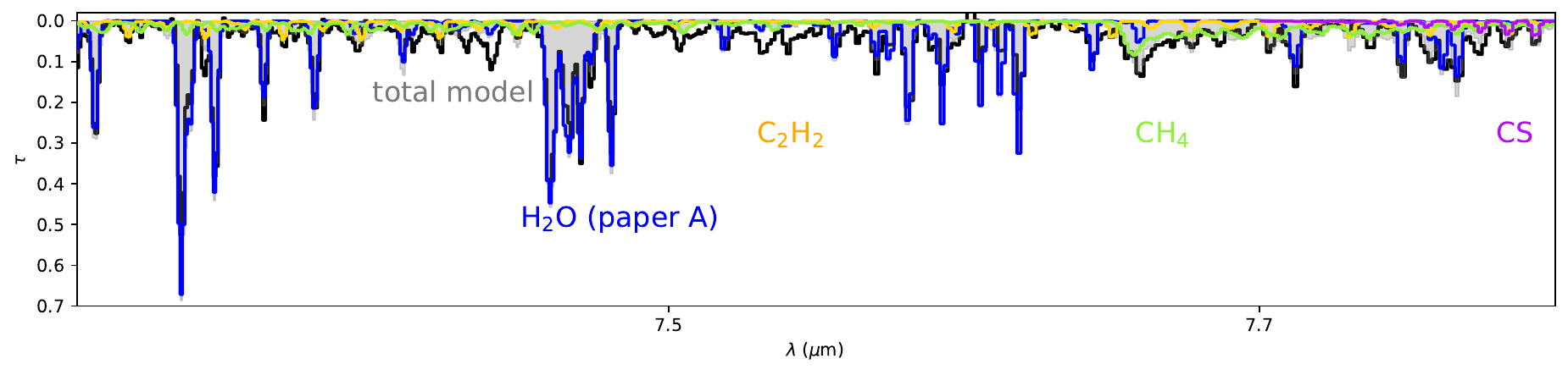}
    \caption{Slab model fits for $\rm{CH_4}$, $\rm{C_2H_2}$, and $\rm{CS}$ in the 7.3 - 7.8 \mum region, fit simultaneously to the residuals of the $\rm{H_2O}$ slab model from paper A. 
    The hydrocarbons were fit with a fixed $\Delta V$ and $v_{\rm{obs}}$. Only the $v_{\rm{obs}}$ for $\rm{CH_4}$ was left free. We indicate the combined model of all four species in this region in gray.}
    \label{fig:CH4_etc_fit}
\end{figure}

\begin{figure}[h!]
    \centering
    \includegraphics[width=\linewidth]{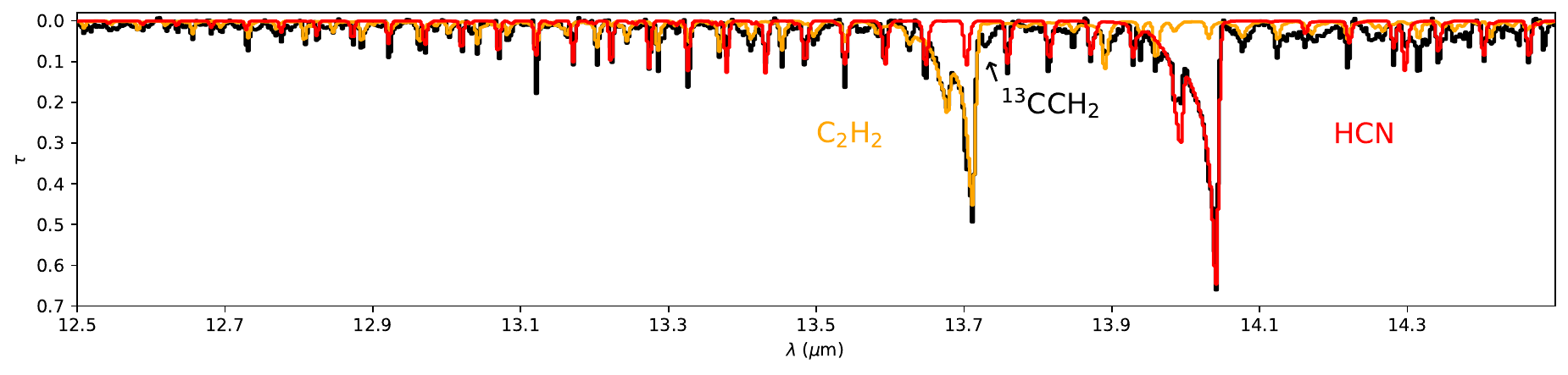}
    \caption{Slab model fits for HCN and  $\rm{C_2H_2}$.  $\rm{C_2H_2}$ was fitted to the residuals of the HCN fit.}
    \label{fig:C2H2_HCN_fit}
\end{figure}

\section{Continuum subtraction and slab model fitting}\label{appendix:slabmodel}
We determined a local baseline on the JWST MIRI spectrum using a guided cubic spline, isolating the gas features from the continuum and solid state features. We then converted to optical depth units using $\tau_{\rm{obs}}=-\ln (F_{\rm{obs}}/F_{\rm{baseline}})$.

We fitted local thermodynamic equilibrium (LTE) slab models in absorption, using the \textit{Molines} code \citep{grant_minds_2023, tabone_rich_2023, francis_joys_2024, van_gelder_jwst_2024}, to a number of absorbing molecular gas bands in order to trace the gas temperatures $T$, column densities $N$, line width $\Delta V$, and velocities $v_{\rm{obs}}$ of the molecular bands. We fit these models to the optical depth spectrum of each feature, using a Bayesian analysis. $\sigma_\tau$ was kept as a free parameter, with a minimum value of 0.01 based on the $1\%$ accuracy of fringe correction achieved by \cite{argyriou_jwst_2023} and \cite{crouzet_extended_2025}. The MCMC fitting then returns the value for $\sigma_\tau$ that brings $\chi^2_{red}$ closest to unity. Such a method was also used in e.g. \cite{kaeufer_bayesian_2024}. We assume a uniform prior over the model parameters $N, T, \Delta V$, $v_{\rm{obs}}$, and $\sigma_\tau$.  Full details can be found in paper A.

In this work, we fit slab models to HCN, $\rm{C_2H_2}$, $\rm{C_4H_2}$, $\rm{CH_3}$, $\rm{CH_4}$, and $\rm{CS}$. If multiple spectral bands overlapped, we typically first fit the most dominant band, and then fit the other band(s) to the residuals of the first fit. This approach was deemed sufficient when individual absorption lines are narrow and do not overlap, as is the case for the overlapping $\rm{HCN}$ and $\rm{C_2H_2}$ bands at 12.5-14.5 $\rm{\mu m}$, and the $\rm{CH_3}$ and $\rm{C_4H_2}$, and $\rm{C_6H_2}$ bands that were fit to the residuals of the 15.5-17 $\rm{\mu m}$ \water band (paper A). 

In the 7.3-7.8 \mum region, multiple molecular bands significantly overlapped, hindering us from fitting them separately. Therefore we fit bands of $\rm{C_2H_2}$, $\rm{CH_4}$, and CS, simultaneously to the residuals of the fit to the \water band (paper A), see Figure\,\ref{fig:CH4_etc_fit}. We masked out any negative optical depth in the residual spectra, to correct for over-subtraction. To decrease the degeneracy introduced by a multi-component fit, we fixed $\Delta V$ and $v_{\rm{obs}}$. However, for further interpretation we deemed it important to fit $v_{\rm{obs}}$ for $\rm{CH_4}$, which was kept free. 

\section{Carbon chain chemistry}\label{appendix:carbonchains}
In the JWST MIRI spectrum, we detect $\rm{C_6H_2}$, a long aliphatic carbon chain species with one hydrogen atom at either end.
The $\rm{C_6H_2}$ 16.1 \mum transition has previously only been detected towards a planetary nebula \citep[e.g. ][]{cernicharo_infrared_2001}, in the Large Magellanic Cloud \citep{bernard-salas_spitzer_2006}, and in an AGN \citep[][]{garcia-bernete_abundant_2026}. The only detections of $\rm{C_6H_2}$ in a star-forming region are of the cumulene isomer (where both hydrogen atoms are on the same end of the carbon chain) in the cold starless core TMC-1 \citep[at 18.8 and 21.5 GHz, e.g. ][]{langer_first_1997} and in a low-mass star-forming region of L1527 \citep[at 21.5 GHz, e.g. ][]{sakai_detection_2007}. 
\cite{shindo_ir_2001} find that for polyynes (carbon chains with the formula $\rm{C_{2N}H_2}$), the peak wavelength of the bending mode band converges to $\sim 621\,\rm{cm^{-1}}$ (16.1 $\rm{\mu m}$) for $\rm{N\geq 3}$. Therefore, we are not able to distinguish between a detection of $\rm{C_6H_2}$ and for example $\rm{C_8H_2}$. 

Longer carbon-chains such as $\rm{C_4H_2}$ and $\rm{C_6H_2}$ can efficiently be formed through gas-phase reactions with $\rm{C_2H}$ in photodominated environments, through reactions such as 
\begin{equation}
    \rm{C_2H + C_2H_2 \rightarrow C_4H_2 + H}
\end{equation}
which is rapid at temperatures above 300\,K \citep{baulch_evaluated_1992}, but can also occur abundantly at lower temperatures \citep{herbst_rate_1997}. This reaction can then be followed up by 
\begin{equation}
    \rm{C_2H + C_4H_2 \rightarrow C_6H_2 + H}
\end{equation}
\citep{tanzawa_thermal_1980}.

In the outflow cavity, $\rm{C_4H_2}$ and $\rm{C_6H_2}$ might be vulnerable to photodissociation. We suspect the hydrocarbon chains could survive if these subsequent chemical reactions occur slightly deeper into the outflow cavity walls, where $\rm{C_2H}$ is still abundant but the resulting hydrocarbon chains are not immediately photodissociated. $\rm{C_4H_2}$ is roughly at rest with the source velocity, which is in line with it being formed in the outflow cavity walls rather than the disk wind.

An alternative route to create the higher order polyynes is through
\begin{equation}
    \rm{C_4H + C_2H_2 \rightarrow C_6H_2 + H}
\end{equation} \citep{sun_formation_2015}. This reaction requires $\rm{C_4H_2}$ to photodissociate into $\rm{C_4H}$, which should happen abundantly in the outflow cavity. This would imply that the photodissociation of lower-order polyynes would actually enable higher-order polyynes to form. Chemical modeling is required to fully understand the influence of the UV field on the carbon-chain formation in protostellar environments. 

HOPS-68 is a known variable \citep{park_quantifying_2021}, further complicating the effect of the UV field on the formation of hydrocarbon chains. \cite{poteet_anomalous_2013} suggests that HOPS-68 recently underwent an outburst, based on the anomalous $\rm{CO_2}$ ice band of HOPS-68. We suggest the increased UV field during an active phase might have caused a higher abundance of e.g. $\rm{C_2H}$ and $\rm{C_4H}$, and during the quiescent phase these species would allow the carbon-chains to form without being photodissociated. 

\end{document}